\documentclass[12pt]{article}

\ifx\pdfoutput\undefined
\usepackage[dvips,bookmarks]{hyperref}
\else
\usepackage{hyperref}
\fi
\hypersetup{colorlinks=false,bookmarksopen,bookmarksnumbered,citecolor=blue,
   pdfstartview=FitH}

\usepackage[pdftex]{graphicx}	
\usepackage{latexsym}

\usepackage{amssymb,amsfonts,amsmath}
\usepackage{graphicx} 
\usepackage{indentfirst}

 \usepackage{bbm}

\parskip=\medskipamount

\newcommand {\cD}{{\cal D}}
\newcommand {\cE}{{\cal E}}

\newcommand {\cN}{{\cal N}}
\newcommand {\cO}{{\cal O}}

\def\a{\alpha}
\def \bi{\bibitem}

\def\b{\beta}

\def\d{\delta}

\def\g{\gamma}
\def\G{\Gamma}

\def\l{\lambda}

\def\n{\nu}
\def\o{\omega}
\def\p{\pi}
\def\q{\theta}

\def\s{\sigma}

\def\D{\Delta}

\def\J{\Psi}
\def\L{\Lambda}
\def\O{\Omega}

\def\rd{{\rm d}}
\def\ri{{\rm i}}
\def\re{{\rm e}}

\newcommand{\ad}{{\dot{\alpha}}}                           %new
\newcommand{\bd}{{\dot{\beta}}}                            %new
\newcommand{\ve}{\varepsilon}                            %new
\newcommand{\cDB}{{\bar\cD}}                            %new

\newcommand{\pa}{\partial}                           %new
\newcommand{\hf}{\frac12}
\newcommand{\vf}{\varphi}
\newcommand{\be}{\begin{equation}}
\newcommand{\ee}{\end{equation}}
\newcommand{\bea}{\begin{eqnarray}}
\newcommand{\eea}{\end{eqnarray}}
\newcommand{\non}{\nonumber}

\newcommand{\dsC}{{\mathbb C}}

\def\double #1{#1{\hbox{\kern-2pt $#1$}}}

\newcommand{\gd}{{\dot\g}}
\newcommand{\dd}{{\dot\d}}

\newcommand{\sba}{{\bar{\s}}}

\newif\ifdtup

\newcommand{\bsubeq}{\begin{subequations}}
\newcommand{\esubeq}{\end{subequations}}
\newcommand{\fF}{\mathfrak F}

\newcommand{\fS}{\mathfrak S}

\numberwithin{equation}{section}

\begin{document}
%%%%%%%%%%%%%%%%
%%%%%%%%%%%%%%%%
\begin{titlepage}
\begin{flushright}
January, 2013\\
\end{flushright}
\vspace{5mm}

\begin{center}
{\Large \bf 
Duality rotations in supersymmetric  nonlinear electrodynamics revisited
}\\ 
\end{center}

\begin{center}

{\bf
Sergei M. Kuzenko
} \\
\vspace{5mm}

\footnotesize{
{\it School of Physics M013, The University of Western Australia\\
35 Stirling Highway, Crawley W.A. 6009, Australia}}  
\vspace{2mm}

\end{center}

\begin{abstract}
\baselineskip=14pt
We revisit the U(1) duality-invariant nonlinear models for $\cN=1$ and $\cN=2$ vector 
multiplets coupled to off-shell supergravities.
For such theories we develop new formulations which make use of auxiliary chiral superfields
(spinor in the $\cN=1$ case and scalar for $\cN=2$) and are characterized by  the remarkable property that 
U(1) duality invariance is equivalent to the manifest U(1) invariance of the self-interaction.
Our construction is inspired by the non-supersymmetric approach that was proposed by Ivanov 
and Zupnik a decade ago and recently re-discovered in the form of twisted self-duality. 
\end{abstract}

\vfill

\vfill
\end{titlepage}

\newpage
\renewcommand{\thefootnote}{\arabic{footnote}}
\setcounter{footnote}{0}

%
%\tableofcontents{}
%\vspace{1cm}
%\bigskip\hrule

%%%%%%%%%%%%%%%%%%%%%%%%%%%%%%%%%%%%%%%%%%%%%%%%
%%%%%%%%%%%%%%%%%%%%%%%%%%%%%%%%%%%%%%%%%%%%%%%%
%%%%%%%%%%%%%%%%%%%%%%%%%%%%%%%%%%%%%%%%%%%%%%%%

\section{Introduction}
\setcounter{equation}{0}

Motivated by patterns of duality in extended supergravity theories \cite{FSZ,CJ}
(for a recent comprehensive review, see \cite{AFZ}), and also extending the famous 1981 work 
by Gaillard and Zumino \cite{GZ1}, the general theory of duality-invariant
models for nonlinear electrodynamics in four dimensions was developed in the mid-1990s
\cite{GR1,GR2,GZ2,GZ3}. 
Given such a model described by a Lorentz invariant Lagrangian $L(F_{ab})$, 
with $F_{ab}$ the electromagnetic field strength, 
the condition for invariance  under U(1) duality rotations\footnote{For early approaches to electromagnetic
duality rotations see \cite{Schrodinger,DeserT}.
For alternative formulations 
of duality symmetric actions see \cite{SS,PST1,PST2} and references therein.}
\bea
\d F_{ab} = \l G_{ab}~, \qquad \d G_{ab} = -\l F_{ab}
\label{1.1}
\eea
proves to be equivalent to the requirement that 
the Lagrangian should obey the equation 
\be
G^{ab} \tilde{G}_{ab} +F^{ab}  \tilde{F}_{ab} = 0~,
\label{GR}
\ee
where
\be
\tilde{G}_{ab} (F):=
\hf \, \ve_{abcd}\, G^{cd}(F) =
2 \, \frac{\pa L(F)}{\pa F^{ab}}~,\qquad
G(F) =  \tilde{F} + \cO(F^3)~.
\label{tilde-g}
\ee
The self-duality equation \eqref{GR}
was originally derived by Gibbons and Rasheed in 1995 \cite{GR1}.
Two years later, it was re-derived by  Gaillard and Zumino  \cite{GZ2} with the aid of their formalism 
developed back in 1981 \cite{GZ1} but originally applied only in the linear case. 
The self-duality equation \eqref{GR} can be reformulated in a form suitable for theories 
with higher derivatives \cite{KT2} (see also \cite{Chemissany:2011yv} for a recent discussion 
with examples). 

As field theories, the models for nonlinear electrodynamics with U(1) duality invariance  
possess  very interesting properties \cite{GZ2,GZ3} reviewed in \cite{KT2} and later in 
\cite{AFZ}. First of all, the energy-momentum tensor is duality-invariant. Secondly, the action 
is automatically invariant under a Legendre transformation, and this is one of the reasons 
why the duality-invariant theories 
may be called self-dual.  Thirdly, although the Lagrangian is not invariant under the duality rotations 
\eqref{1.1}, 
\bea
\d L = \frac{1}{4} \l (G^{ab}\tilde{G}_{ab} - F^{ab}\tilde{F}_{ab})~,
\eea 
the partial derivative $\pa L/\pa g$ with respect to any duality-inert parameter $g$ 
is invariant under \eqref{1.1}. In fact, the duality invariance of the energy-momentum tensor 
is a corollary of this general statement. 
It is worth pointing out that for any solution $L(F_{ab}) $ of the 
self-duality equation and a real parameter $g$, the Lagrangian 
\bea
\hat{L} (F_{ab}):= \frac{1}{g^2} L(g F_{ab}) 
\eea
is also a solution of \eqref{1.1}.

The concept of self-dual nonlinear electrodynamics was generalized  to the cases
of $\cN=1$ and $\cN=2$ rigid supersymmetric theories in \cite{KT1}.
This generalization has turned out to be very useful, since the families of actions 
obtained include all the known models 
for partial breaking of supersymmetry based 
on the use of a vector Goldstone multiplet.
In particular, the $\cN=1$ supersymmetric Born-Infeld
action \cite{CF}, which  is a Goldstone multiplet 
action for partial supersymmetry 
breakdown $\cN=2 \to \cN=1$ \cite{BG,RT} is, at the same time,   a solution 
to the $\cN=1$  self-duality equation 
\cite{KT2,KT1}.  Furthermore, the model for partial 
breaking of supersymmetry $\cN=4 \to \cN=2$  \cite{BIK}, which nowadays is identified
with the $\cN=2$ supersymmetric Born-Infeld action, 
was first constructed in \cite{KT2}
as a unique solution to  the $\cN=2$ self-duality equation 
possessing a nonlinearly realized central charge 
symmetry. 
The models for self-dual nonlinear supersymmetric electrodynamics \cite{KT2,KT1}  were generalized  to 
$\cN=1$ supergravity in \cite{KMcC} and recently to $\cN=2$ supergravity \cite{K_N=2}.

The self-duality equation \eqref{GR} is a nonlinear differential equation on the Lagrangian, 
and thus its general solutions are difficult to construct explicitly. 
The most famous exact solution of \eqref{GR}  is the Born-Infeld 
Lagrangian \cite{BI}
\bea
L_{\rm BI} (F_{ab}) &=& \frac{1}{g^2} \Big\{
1 - \sqrt{- \det (\eta_{ab} + g F_{ab} )} 
\Big\}~,
\label{BI}
\eea
with $g$ the coupling constant. 

A decade ago, Ivanov and Zupnik \cite{IZ1,IZ2} proposed a reformulation of nonlinear 
electrodynamics, $L(F_{ab}) \to \widetilde{L}(F_{ab} , V_{ab})$, which makes use 
of an auxiliary bivector $V_{ab}=-V_{ba}$, the latter being  equivalent to a pair of symmetric spinors,
$V_{\a\b}= V_{\b\a}$ and its conjugate $\bar  V_{\ad\bd}$.
The new Lagrangian $\widetilde{L}$ is at most quadratic with respect to
the electromagnetic field strength $F_{ab}$, while the self-interaction is described 
by a nonlinear function of the auxiliary variables, $L_{\rm int} (V_{ab})$,
\bea
 \widetilde{L}(F_{ab} , V_{ab}) = \frac{1}{4} F^{ab}F_{ab} +\hf V^{ab}V_{ab} 
 - V^{ab}F_{ab} + L_{\rm int} (V_{ab})~.
\eea 
The original theory  $L(F_{ab})$ is obtained from 
$\widetilde{L}(F_{ab} , V_{ab})$ by integrating out the auxiliary variables. 
In terms of $\widetilde{L}(F_{ab} , V_{ab})$, the condition of U(1) duality invariance 
was shown \cite{IZ1,IZ2}  to be equivalent to the requirement that the self-interaction
\bea
L_{\rm int} (V_{ab}) = L_{\rm int} (\n, \bar \n)~, \qquad \n:=V^{\a\b}V_{\a\b}
\eea
is invariant under linear U(1)  transformations $\n \to \re^{\ri \vf} \n$, with $\vf \in \mathbb R$,
and thus 
\bea
L_{\rm int} (\n, \bar \n)= f (\n \bar \n)~,
\label{1.5}
\eea
where $f$ is a real function of one real variable. As a result, the Ivanov-Zupnik 
formulation allows one to generate, in principle, all solutions of the self-duality equation. 
At first sight, this approach appears somewhat mysterious. However its origin becomes more transparent 
if we recall some general features of all solutions of \eqref{GR} discussed in \cite{GR1,KT2}.

First of all, it is worth recalling another useful representation for the Lagrangian $L(F_{ab})$ and 
 for the self-duality equation, following \cite{KT2}.
Since in four dimensions
the electromagnetic field has only two independent invariants, 
\bea
\a = \frac{1}{4} \, F^{ab} F_{ab}~, \qquad \quad
\b = \frac{1}{4} \, F^{ab} \tilde{F}_{ab} ~,
\eea
the Lagrangian $L(F_{ab})$ can be considered as a
real function of one complex variable
\bea
L(F_{ab})= L(\o , \bar{\o} )~, \qquad \quad
\o = \a + {\rm i} \, \b~.
\label{omega}
\eea
The theory is parity invariant iff $L(\o , \bar{\o} )
= L( \bar{\o}, \o )$. If the theory is duality invariant, then $L(\o , \bar{\o} )$
can be shown \cite{KT2} to have the form 
\bea
L (\o , \bar{\o} )= -\hf \, \big( \o + \bar{\o} \big) + \o  \bar{\o} \, \L (\o, \bar{\o} )~,
\qquad \quad \L(\o , \bar{\o} ) = {\rm const} ~+~ \cO (|\o|)~,
\label{1.11}
\eea
where the interaction $ \L (\o, \bar{\o} )$ is a real analytic function.
The self-duality equation \eqref{GR} turns into 
\bea
{\rm Im}\;  \left\{ \pa_\o (\o \, \L) 
- \bar{\o}\,
\Big( \pa_\o (\o \, \L )   \Big)^2 \right\} = 0~,
\label{GR2}
\eea
with $\pa_\o =\pa / \pa \o$. 
For the Born-Infeld Lagrangian \eqref{BI}, we have 
\bea
L_{\rm BI} (\o , \bar{\o} ) &=& 
  \frac{1}{g^2} \left\{ 
1 - \sqrt{1 + g^2 (\o + \bar \o )
+\frac{1 }{ 4}g^4 (\o - \bar \o )^2 } 
\right\}~, \non \\
\L_{\rm BI}  (\o , \bar{\o} )&=&
\frac{g^2 }{ 1 +
\hf g^2(\o + \bar \o )
+ \sqrt{1 + g^2 (\o + \bar \o )
+\frac{1}{4}g^4 (\o - \bar \o )^2 }}~.
\label{bi-lag}
\eea
It is a simple exercise to check that $\L_{\rm BI}  $ is a solution of \eqref{GR2}.

Now, we reproduce {\it verbatim} a paragraph from section 2 in \cite{KT2} (a similar discussion appeared 
earlier in \cite{GR1}). 
\begin{quote}
In perturbation theory one looks for a parity invariant solution
of the self-duality equation by considering the Ansatz
\bea
\L (\o , \bar \o ) ~=~ \sum_{n=0}^{\infty} ~
\sum_{p+q =n} C_{p,q} \; \o^p {\bar \o}^q~,
\qquad \quad C_{p,q}=C_{q,p} ~\in~ {\mathbb R}~,
\label{self-interaction}
\eea
where $n=p+q$ is
the level of the coefficient $C_{p,q}$.
It turns out that for odd level
the self-duality equation uniquely expresses all
coefficients recursively.
If, however, the level is even, the self-duality equation
uniquely fixes the level-$n$ coefficients
$C_{p,q}$ with $p \neq q$ through those at lower levels,
while $C_{r,r}$ remain undetermined.
This means that a general solution of the self-duality equation
involves an arbitrary real analytic function of
one real argument, $f(\o \bar \o )$.
\end{quote}
Given a real analytic function 
\bea
f(\o \bar \o )=
\sum_{r=0}^{\infty} C_{r,r} \; (\o \bar \o)^r~,
\label{1.16}
\eea
the self-duality equation \eqref{GR2} uniquely determines the entire self-interaction
\eqref{self-interaction}. This means that there exists a one-to-one map
$\p: f(\o \bar \o )\to \L (\o , \bar \o )$, where  $ \L (\o , \bar \o )$ corresponds to a duality-invariant theory. 
In other words, the duality-invariant theories
can be formulated in terms of the function $f(\o \bar \o )$. This is actually the same function 
which appears within the Ivanov-Zupnik approach, eq. \eqref{1.5}. Their approach is essentially 
a scheme to formulate self-dual theories in terms of such a real function, $f_{\rm IZ}(x) $, uniquely related
to $f(x) $ in \eqref{1.16}.
 
Recently, there has been a revival of interest in the duality-invariant dynamical systems
\cite{BN,CKR,Chemissany:2011yv,BCFKR} inspired by the desire to achieve a better understanding
of   the UV properties of extended supergravity theories. 
The authors of \cite{BN,CKR,Chemissany:2011yv} 
have put forward the so-called ``twisted self-duality constraint'' 
as a systematic procedure to generate duality-invariant theories. 
However, it has been demonstrated \cite{IZ3} that the non-supersymmetric construction of  
\cite{BN,CKR,Chemissany:2011yv} naturally originates within the more general approach 
previously developed in \cite{IZ1,IZ2}. Specifically, the twisted self-duality constraint 
correspond to an equation of motion in the approach of \cite{IZ1,IZ2}. 

The authors of \cite{BCFKR} studied perturbative solutions of the $\cN=2$ supersymmetric
self-duality equation \cite{KT1} by combining the perturbative analysis of \cite{KT2} with the idea 
of twisted self-duality. In the present paper, we give $\cN=1$ and $\cN=2$ locally 
supersymmetric extensions of the Ivanov-Zupnik approach. In the rigid supersymmetric limit,
our results provide an off-shell extension of the approach pursued in  \cite{BCFKR}. 

This paper is organized as follows. In section 2 we give a brief summary of self-dual models
for $\cN=1$ supersymmetric electrodynamics coupled to supergravity. Here two off-shell 
realizations for $\cN=1$ supergravity are used: the old minimal formulation \cite{WZ,old} 
and the new minimal formulation \cite{new}. In section 3 we develop a novel description of 
the $\cN=1$ supersymmetric duality-invariant theories \cite{KT1,KT2,KMcC}
that makes use of an auxiliary covariantly chiral  spinor superfield and its conjugate. 
Section 4 is devoted to  a novel description of 
the $\cN=2$ supersymmetric duality-invariant theories \cite{KT1,KT2,K_N=2} 
that employs an auxiliary covariantly chiral  scalar superfield and its conjugate. 
A few concluding comments are given in section 5. The main body of the paper is accompanied by 
two technical appendices devoted to aspects of the superspace differential geometry of $\cN=1$ and 
$\cN=2$ supergravities. 

\section{Duality rotations in $\cN=1$ supersymmetric nonlinear electrodynamics}

Unless otherwise specified, in this and the next sections we use the old minimal formulation 
for $\cN=1$ supergravity. Our notation and conventions mostly follow \cite{BK}
(which are similar to those adopted in \cite{WB}) with the only exception 
that we use different symbols for the full superspace and for the chiral integration measures.  
A summary concerning the Wess-Zumino superspace geometry is given in Appendix A. 

Consider a theory of an  Abelian $\cN = 1$
vector multiplet in curved superspace generated by an action
$S[W , {\bar W}]$. The covariantly  chiral spinor field strength $W_\a$
 and its conjugate ${\bar W}_\ad$ are defined as 
\bea
W_\a = -\frac{1}{4}\,  (\bar \cD^2 -4R) 
 \cD_\a  V~, \qquad \quad
{\bar W}_\ad = -\frac{1}{4}\,   ( \cD^2 -4\bar R) \bar \cD_\ad  V ~,
\label{eq:w-bar-w}
\eea
 in terms of a real unconstrained prepotential $V$.
The field strengths $W_\a$ and ${\bar W}_\ad$  obey
the Bianchi identity
\bea
\cD^\a W_\a = \bar \cD_\ad {\bar W}^\ad~.
\label{eq:bianchi}
\eea
In many cases $S[W , {\bar W}] $ can unambiguously be defined
as a functional of an {\it unrestricted} covariantly 
 chiral superfield $W_\a$ and its conjugate ${\bar W}_\ad$.\footnote{As pointed out in \cite{KT1,KT2},
this is always possible if $S[W , {\bar W}]$
does not involve the combination $\cD^\a W_\a $ as an
independent variable.} Then, defining 
\bea
{\rm i}\,M_\a := 2\, \frac{\d }{\d W^\a}\,S[W , {\bar W}]~,
\label{2.3M}
\eea
the equation of motion for $V $ is 
\bea
\cD^\a M_\a = \cDB_\ad {\bar M}^\ad~.
\label{eq:eom}
\eea
Here the variational derivative $\d S/\d W^\a $ is defined by 
\bea
\d S =  \int \rd^4 x \,{\rm d}^2 \q \,\cE\, \d W^\a \frac{\d S}{\d W^\a}~+~{\rm c.c.}~,
\eea
where $\cE$ denotes the chiral integration measure, and $W_\a$ is
assumed to be an unrestricted covariantly chiral spinor.

Since the Bianchi identity (\ref{eq:bianchi}) and the equation of
motion (\ref{eq:eom}) have the same functional form, one may
consider U(1) duality rotations
\bea
\d W_\a = \l M_\a ~, 
\qquad \d M_\a = - \l W_\a~,
\label{DualRot}
\eea
with $\l$ a constant parameter. The condition for duality invariance is the self-duality equation 
\bea
{\rm Im} \int \rd^4 x \rd^2 \q  \,\cE \Big\{ W^2 +M^2 \Big\} =0~,
\label{2.6}
\eea
in which $W_\a$ is chosen to be an unrestricted covariantly chiral spinor. 

For any vector multiplet model  $S[W , {\bar W}]$, one can develop a dual formulation.  
This is achieved by introducing the auxiliary action
\bea
S[{W} , {{\bar W}}, W_{\rm D}, \bar W_{\rm D}] =
-\frac{{\rm i}}{ 2} \int \rd^4 x \rd^2 \q  \,\cE  \,{W}^\a W_{{\rm D}\, \a} +{\rm c.c.}
+S[{W},{\bar W} ]~,
\label{n=1da}
\eea
where ${W}_\a$ is now an unrestricted covariantly chiral spinor
superfield, and $W_{{\rm D}\, \a}$ the dual field strength
\bea
W_{{\rm D}\,\a} = -\frac{1}{4}\, ({\bar \cD}^2 -4R)\cD_\a \, V_{\rm D}~,\qquad \bar V_{\rm D}= V_{\rm D}~,
\eea
with $V_{\rm D}$ a dual gauge prepotential. 
This model is equivalent to the original model, since
the equation of motion for $V_{\rm D}$ implies that
${W}$ satisfies the Bianchi identity \eqref{eq:bianchi}
and the action  (\ref{n=1da}) reduces to $S[W , {\bar W}]$.
On the other hand, under quite general conditions on the structure of $S[{W},{\bar W} ]$, 
one can integrate out from $S[{W} , {{\bar W}}, W_{\rm D}, \bar W_{\rm D}] $ the auxiliary 
variables $W_\a$ and $\bar W_\ad$ and  end up with 
a dual action $S_{\rm D}[W_{\rm D} , {\bar W}_{\rm D}]$. 
By construction the models  $S [W , {\bar W} ]$ and  $S_{\rm D}[W , {\bar W}]$ are related 
to each other by a Legendre transformation. 

If $S[W , {\bar W}]$ is a solution of the self-duality equation \eqref{2.6}, then the dual action 
has the same functional form as the original action, 
\bea
S_{\rm D}[W , {\bar W}] = S[W , {\bar W}]~.
\label{Legendre}
\eea
Therefore the theory is self-dual under the superfield Legendre transformation. 

The most general self-dual model with no more than  two derivatives at the component level 
was constructed in the rigid supersymmetric case in  \cite{KT1} and extended to supergravity 
a few years later in \cite{KMcC}. The corresponding action has the form 
\bea
S_{\rm SED}=  \frac{1}{4}\int \rd^4 x \rd^2 \q  \,\cE \,  W^2 +{\rm c.c.} 
+ \frac{1}{4} \int \rd^4 x \rd^2 \q \rd^2\bar \q \,E \,W^2 \bar W^2 \L(u, \bar u)~,
\label{MasterModel}
\eea
where 
\bea
u:= \frac{1}{8} (\cD^2 -4\bar R) W^2~.
\eea
For this model the self-duality equation \eqref{2.6}  amounts to
\bea
{\rm Im} \int  \rd^4 x \rd^2 \q \rd^2\bar \q \,E \, W^2 \bar W^2 \Big\{ \G
- \bar{u}\, \G^2
\Big\} = 0~, \qquad \quad
\G  := \pa_u (u \, \L) ~.
\label{2.9SDE}
\eea
In this equation the covariantly chiral spinor $W_\a$ has to be completely arbitrary, 
and therefore we conclude that 
\bea
{\rm Im} \Big\{ \G
- \bar{u}\, \G^2
\Big\} = 0~.
\eea
The component structure of the theory \eqref{MasterModel} was studied in \cite{KMcC2}.
In the rigid supersymmetric case, the bosonic sector of  \eqref{MasterModel}
was originally analyzed in \cite{KT2}. Upon switching off the auxiliary field of the vector multiplet, 
$D=0$, which always is a solution of the corresponding equation of motion, 
the bosonic action reduced to that describing self-dual nonlinear electrodynamics. 
The self-interaction $\L$, which determines the bosonic Lagrangian \eqref{1.11},  coincides with that 
appearing in the supersymmetric action  \eqref{MasterModel}, see \cite{KT2,KMcC2} for more details. 

The supercurrent multiplet corresponding to the self-dual theory \eqref{MasterModel}
was computed in \cite{KMcC} and shown to be duality-invariant. 

The action \eqref{MasterModel} describes supersymmetric nonlinear electrodynamics 
in old minimal supergravity. The model can also be coupled to new minimal supergravity \cite{new}
or to non-minimal supergravity \cite{non-min,SG}. 
Here we recall, following \cite{KMcC},
 how this is achieved in the case of new minimal supergravity. 

As is known,  each off-shell formulation for $\cN=1$ supergravity 
can be realized as a super-Weyl invariant coupling of  old minimal supergravity  
to certain compensator(s)  \cite{BK,FGKV,KU}.
Super-Weyl transformations \cite{HT} are local scale and U(1) transformations 
of the covariant derivatives of the form 
\bea
\d_\s \cD_\a =  ({\bar \s} -  \hf  \s  )
\cD_\a + (\cD^\b \s)  M_{\a \b}  ~, \qquad
\d_\s \cDB_\ad = ( \s-  \hf  {\bar \s}   )
\cDB_\ad +  (\cDB^\bd {\bar \s}) {\bar M}_{\bd\ad} ~,
\label{superweyl}
\eea
with $\s $ an arbitrary  covariantly chiral scalar parameter,
$\cDB_\ad \s=0$. The compensator for new minimal supergravity 
is a real covariantly linear scalar $\mathbb L$, 
\bea
(\bar \cD^2 - 4 { R})\, {\mathbb L} = 0~, \qquad \bar {\mathbb L} ={\mathbb L}~,
\label{linear}
\eea
which is required to be nowhere vanishing. 
Its super-Weyl transformation law is 
\bea
\d_\s {\mathbb L} =(\s + \bar \s ) \,{\mathbb L}~.
\eea
Recalling the super-Weyl transformation of $W_\a$, 
\bea
\d_\s W_\a = \frac{3}{2} \s \, W_\a ~,
\label{2.18W}
\eea
one may see that the following combination 
\bea
(\cD^2 - 4 {\bar R}) \Big( \frac{W^2 }{ {\mathbb L}^2 } \Big)
\eea
is super-Weyl invariant. 
This implies that the vector-multiplet action \cite{KMcC}
\bea
S[W,{\bar W},{\mathbb L}] &=&
\frac{1}{4}\int  \rd^4 x \rd^2 \q  \,\cE \, W^2 +{\rm c.c.}
\non \\
&&{}\quad + \frac14\,  \int \rd^4 x \rd^2 \q \rd^2\bar \q \,E \,
\frac{W^2\,{\bar W}^2}{{\mathbb L}^2}\,
\L\!\left(\frac{u}{{\mathbb L}^2},
\frac{\bar u}{{\mathbb L}^2}\right)~,
\label{SED-NSG}
\eea
is super-Weyl invariant. Moreover, it is not difficult to check that $S[W,{\bar W},{\mathbb L}] $
 solves the self-duality equation \eqref{2.6}. The action \eqref{SED-NSG} described self-dual 
 supersymmetric electrodynamics coupled to new minimal supergravity.  

\section{New realization}

We now turn to presenting a new formulation for the self-dual models of the $\cN=1$ 
vector multiplet described in the previous  section. This representation is inspired 
by the  non-supersymmetric construction of \cite{IZ1,IZ2}.

\subsection{General setup}
Consider an auxiliary action of the form
\bea
S[W,\bar W, \eta, \bar \eta]= \int \rd^4 x \rd^2 \q  \,\cE \Big\{ \eta W -\hf \eta^2 - \frac{1}{4} W^2\Big\} 
+{\rm c.c.}  + \fS_{\rm int} [\eta, \bar \eta]~.
\label{3.111}
\eea
Here the spinor superfield $\eta_\a$ 
is constrained to be covariantly chiral, $\bar \cD_\bd \eta_\a =0$, 
but otherwise it is completely arbitrary. 
By definition, the second term on the right, $ \fS_{\rm int} [\eta, \bar \eta]$,
contains cubic, quartic and higher powers of $\eta_\a$ and its conjugate. 

The above model is equivalent to a theory with action 
\bea
S[W,\bar W] =    \frac{1}{4}\int \rd^4 x \rd^2 \q  \,\cE  \, W^2 
+{\rm c.c.}  + S_{\rm int} [W, \bar W]~,
\label{VMaction}
\eea
describing the dynamics
of the vector multiplet. Indeed, the equation of motion for $\eta^\a$ is 
\bea
W_\a = \eta_\a - \frac{\d}{\d \eta^\a} \fS_{\rm int} [\eta, \bar \eta]~.
\label{EoM}
\eea
In perturbation theory, this equation can be used to express $\eta_\a$ as a functional 
of $W_\a$ and its conjugate, $\eta_\a = \J_\a[W, \bar W]$. Plugging this functional and its conjugate
into the action \eqref{3.111}, we end up with some  vector-multiplet model 
of the form \eqref{VMaction}.
As a result, we have two equivalent realizations of the same theory, in terms of the action 
\eqref{3.111} or in terms of $S[W,\bar W]$.

Suppose $S[W,\bar W]$ is a solution of the self-duality equation \eqref{2.6}.
We need to understand what the implications of self-duality 
are on the structure of \eqref{3.111}. 
Let us compute $M_\a$ by using the two actions $S[W,\bar W, \eta, \bar \eta]$ and  $S[W,\bar W]$:
\begin{subequations}
\bea
{\rm i}\,M_\a := 2\, \frac{\d }{\d W^\a}\,S &=& 2\eta_\a -W_\a
\label{3.3M} \\
&=& W_\a + 2  \frac{\d}{\d W^\a} S_{\rm int} [W, \bar W]~.
\label{3.3Mb} 
\eea
\end{subequations}
Now, if we make use of the equation of motion for $\eta$, eq. \eqref{EoM}, 
the self-duality equation \eqref{2.6} turns into
\bea
{\rm Im} \int \rd^4 x \rd^2 \q  \,\cE \,\eta^\a \frac{\d}{\d \eta^\a} \fS_{\rm int} [\eta, \bar \eta] =0~.
\eea
This  condition means that the self-interaction $\fS_{\rm int} [\eta, \bar \eta] $ is invariant 
under rigid U(1) phase transformations of $\eta_\a$ and its conjugate,
\bea
\fS_{\rm int} [ \re^{\ri \vf}  \eta, \re^{-\ri \vf} \bar \eta] = \fS_{\rm int} [\eta, \bar \eta] ~, 
\qquad \vf \in {\mathbb R}~.
\eea
The duality rotation \eqref{DualRot} acts on 
the chiral spinor $\eta_\a =\hf (W_\a+\ri M_\a)$, eq. \eqref{3.3M},  as
\bea
\d \eta_\a = - \ri \l \eta_\a~.
\label{DualTransformationL}
\eea

${}$From \eqref{3.3M} and \eqref{3.3Mb} we have
\bea
\eta_\a = W_\a +  \frac{\d}{\d W^\a} S_{\rm int} [W, \bar W]~.
\eea
This relation allows us to express $W_\a$ as a functional of $\eta_\a$ and its conjugate, 
$W_\a = W_\a[\eta, \bar \eta]$, and then reconstruct the self-interaction 
$ \fS_{\rm int} [\eta, \bar \eta]$ starting from the action \eqref{VMaction}.

Given a manifestly U(1) invariant self-interaction $\fS_{\rm int} [\eta, \bar \eta] $, 
we can construct a duality-invariant theory  $S[W,\bar W]$ by starting for the action \eqref{3.111}
and then integrating out the auxiliary variables $\eta_\a$ and $\bar \eta_\ad$.
 
 Suppose  that $S[W,\bar W]$ is 
self-dual under Legendre transformation, eq. \eqref{Legendre}.  In terms of the auxiliary action 
\eqref{3.111} this condition proves to be equivalent to 
\bea
\fS_{\rm int} [ \ri  \, \eta, -\ri \, \bar \eta] = \fS_{\rm int} [\eta, \bar \eta] ~.
\eea
We see that the self-duality under the superfield Legendre transformation is equivalent to the fact that 
the self-interaction $\fS_{\rm int} [\eta, \bar \eta]$ is ${\mathbb Z}_4$ invariant.  

In general, self-duality under a Legendre transformation is known to be a pretty mild condition  
\cite{GKPR}.

\subsection{Two-derivative  models}
We now turn to duality-invariant supersymmetric theories with 
at most two derivatives at the component level. 
Consider an auxiliary action of the form 
\bea
S[W,\bar W, \eta, \bar \eta]&=& \int{\rm d}^4 x \rd^2\q\,\cE \Big\{ \eta W -\hf \eta^2 - \frac{1}{4} W^2\Big\} 
+{\rm c.c.} \non \\
&& \quad+ \frac{1}{4} \int \rd^4 x \rd^2 \q \rd^2\bar \q \,E \,\eta^2 \bar \eta^2 \fF(v, \bar v)~,
\label{3.1}
\eea
where 
\bea
v:= \frac{1}{8} (\cD^2 -4\bar R) \eta^2~,
\eea
and $\fF(v, \bar v)$ is a real analytic function. 
We would like to integrate out from   \eqref{3.1}
the auxiliary spinor variables $\eta_\a$ and $\bar \eta_\ad$
 in order to bring the action to the form \eqref{MasterModel}.
The equation of motion for $\eta^\a$ is 
\bea
W_\a = \eta_\a \left\{ 1 + \frac{1}{8} (\bar \cD^2 -4R) 
\Big[ {\bar \eta}^2 \Big(\fF  + \frac{1}{8} (\cD^2 -4\bar R) 
\big( \eta^2\, \pa_v \fF \big) \Big) \Big] \right\}~.
\label{3.3}
\eea
Its immediate implications are 
\begin{subequations} \label{3.4}
\bea
\eta W&=& \eta^2 \Big[ 1 + \frac{1}{8} (\bar \cD^2 -4R) 
\Big\{ {\bar \eta}^2 \pa_v (v \fF)\Big\} \Big] ~, \label{3.4a}\\
 W^2 &=& \eta^2 \Big[ 1 + \frac{1}{8} (\bar \cD^2 -4R) 
\Big\{ {\bar \eta}^2 \pa_v (v \fF)\Big\} \Big]^2 ~, \label{3.4b} \\
W^2 \bar W^2 &=& \eta^2 \bar \eta^2 \Big[ 1+ \pa_v (v\bar v \fF)\Big]^2 
\Big[ 1+ \pa_{\bar v} (v\bar v \fF)\Big]^2~.
\label{3.4c}
\eea
\end{subequations}
Eq. \eqref{3.4b} implies that 
\bea
u \approx v [1+\pa_v (v\bar v {\mathfrak F})]^2 ~. 
\label{3.5}
\eea
Here and below, the symbol $\approx$ is used to indicate that the result holds modulo terms 
proportional to $\eta_\a $ and $\bar  \eta_\ad$ (or, equivalently, 
to $W_\a $ and $\bar  W_\ad$).

The identities \eqref{3.4} may be used to derive several integral relations
\begin{subequations}
\bea
 \int \rd^4 x \rd^2 \q  \,\cE \, W^2 &=&  \int \rd^4 x \rd^2 \q  \,\cE\, \eta^2 \non \\
&& -\int \rd^4 x \rd^2 \q \rd^2\bar \q \,E \,\eta^2 \bar \eta^2 \Big\{ \pa_v(v \fF) +\hf 
\bar v [\pa_v (v\fF)]^2 \Big\} ~, \\
 \int \rd^4 x \rd^2 \q  \,\cE \, \eta W &=&  \int \rd^4 x \rd^2 \q \rd^2 \,\cE\, \eta^2
-\hf \int  \rd^4 x \rd^2 \q \rd^2\bar \q \,E \,\eta^2 \bar \eta^2  \pa_v(v\fF) ~.
\eea
\end{subequations}
With the aid of these relations, the action \eqref{3.1} takes the form 
\bea
S[W,\bar W]=  \frac{1}{4}\int \rd^4 x \rd^2 \q  \,\cE \,  W^2 +{\rm c.c.} 
+ \frac{1}{4} \int \rd^4 x \rd^2 \q \rd^2\bar \q \,E \,W^2 \bar W^2 \L(u, \bar u)~,
\label{3.7}
\eea
where we have introduced
\bea
\L(u,\bar u):= \frac{ \fF + \bar v [\pa_v(v\fF)]^2 + v[\pa_{\bar v}(\bar v \fF)]^2 }
{ \big[ 1+ \pa_v (v\bar v \fF)\big]^2 
\big[ 1+ \pa_{\bar v} (v\bar v \fF)\big]^2}~.
\label{3.8}
\eea
The right-hand side of \eqref{3.8} is uniquely determined in terms of $\fF$ and its partial derivatives. 
In order to read off the function in the left-hand side of \eqref{3.8}, we have to know the expression for 
$u$ as a functional of $v$ and $\bar v$ which, in accordance with \eqref{3.4b}, is quite complicated:
\bea
u= \frac{1}{8}  ( \cD^2 -4\bar R) \left(
\eta^2 \Big[ 1 + \frac{1}{8} (\bar \cD^2 -4R) 
\Big\{ {\bar \eta}^2 \pa_v (v \fF)\Big\} \Big]^2 \right)~.
\label{3.14L}
\eea
However, since $\L(u,\bar u)$ appears in the action \eqref{3.8} multiplied by $W^2 \bar W^2$, 
when evaluating $\L(u,\bar u)$ 
we can replace \eqref{3.5} with the ``effective'' relation $u= v [1+\pa_v (v\bar v {\mathfrak F})]^2$.
The latter allows us to express $v$ in terms of $u$ and $\bar u$, and therefore to read off the function 
$\L(u,\bar u)$ in \eqref{3.14L}.

Now we have to learn how to carry out an inverse transformation, that is how to reconstruct 
the self-coupling $\fF(v,\bar v)$ starting from the action \eqref{3.7}.
Making use of the action \eqref{3.1} leads to  \eqref{3.3M}, and hence
\bea
\eta_\a=\hf(W_\a +\ri M_\a)~.
\eea
On the other hand, we may compute $ M_\a$, eq. \eqref{2.3M},  from the action \eqref{3.7}.
\bea
\ri M_\a = W_\a \left\{ 1 - \frac{1}{4} (\bar \cD^2 -4R) 
\Big[ {\bar W}^2 \Big(\L  + \frac{1}{8} (\cD^2 -4\bar R) 
\big( W^2\, \pa_u \L \big) \Big) \Big] \right\}~.
\eea
Combining the two results, we obtain
\bea
\eta_\a = W_\a \left\{ 1 - \frac{1}{8} (\bar \cD^2 -4R) 
\Big[ {\bar W}^2 \Big(\L  + \frac{1}{8} (\cD^2 -4\bar R) 
\big( W^2\, \pa_u \L \big) \Big) \Big] \right\}~.
\label{3.12}
\eea
An important result may be seen by comparing the relations \eqref{3.3} and \eqref{3.12}.
We obtain
\bea
\big[1-\pa_u (u\bar u \L ) \big] \big[1+ \pa_v (v \bar v\fF)\big] \approx 1~.
\label{3.20}
\eea

There are three simple implications of \eqref{3.12}:
\begin{subequations} \label{3.14}
\bea
\eta W&=& W^2 \Big[ 1 + \frac{1}{8} (\bar \cD^2 -4R) 
\Big\{ {\bar W}^2 \pa_u (u \L)\Big\} \Big] ~, \label{3.14a}\\
 \eta^2 &=& W^2 \Big[ 1 - \frac{1}{8} (\bar \cD^2 -4R) 
\Big\{ {\bar W}^2 \pa_u (u \L)\Big\} \Big]^2 ~. 
\label{3.14b} \\
\eta^2 \bar \eta^2 &=& W^2 \bar W^2 \big[ 1- \pa_u (u\bar u \L)\big]^2 
\big[ 1- \pa_{\bar u} (u\bar u \L)\big]^2~.
\label{3.14c}
\eea
\end{subequations}
These results lead to 
\begin{subequations}
\bea
v&\approx & u[ 1- \pa_u (u \bar u \L)]^2~,\\
v[1+ \pa_v(v\bar v \fF)] &\approx &  u[ 1- \pa_u (u \bar u \L)]~.
\eea
\end{subequations}
Due to \eqref{3.20}, the relation  \eqref{3.4b} is equivalent to \eqref{3.14c}. 
The relations obtained allow us to restore the self-interaction
\bea
\fF(v,\bar v)= \frac{ \L - \bar u [\pa_u(u\L)]^2 - u [\pa_{\bar u}(\bar u \L)]^2 }
{
\big[ 1- \pa_u (u\bar u \L)\big]^2 
\big[ 1- \pa_{\bar u} (u\bar u \L)\big]^2
}~.
\label{3.23}
\eea
Starting from the action \eqref{3.7}, it is now trivial  to restore 
the self-interaction $\fF(v,\bar v)$ by making use of eq. \eqref{3.23} in conjunction 
with the effective relation $v= u[ 1- \pa_u (u \bar u \L)]^2$.
It is instructive to compare the functional forms of the transformation 
$\L(u,\bar u) \to \fF(v,\bar v)$, eq.  \eqref{3.23}, and its inverse \eqref{3.8}. 

The last point to analyze is  U(1) duality invariance. Suppose that the action \eqref{3.7}
is a solution of the self-duality equation \eqref{2.6}. Using the above relations, 
one may show that 
\bea
W^2 \bar W^2 (\G - \bar u \G^2) = \eta^2 \bar \eta^2(\fF + v \pa_v \fF) ~.
\eea
Therefore, the  self-duality condition \eqref{2.9SDE} is equivalent to 
\bea
( v \pa_v - \bar v \pa_{\bar v} ) \fF =0\quad \Longleftrightarrow \quad
\fF (v, \bar v) = f(v \bar v)~.
\eea
We conclude  that the self-interaction $\fF (v, \bar v) $ must be invariant under the U(1) 
transformations  \eqref{DualTransformationL}.

The model \eqref{3.1} can naturally be coupled to new minimal supergravity. 
For this we postulate the super-Weyl transformation of $\eta_\a$ to be (compare with \eqref{2.18W})
\bea
\d_\s \eta_\a = \frac{3}{2} \s \eta_\a
\eea
and replace the action \eqref{3.1} with
\bea
S[W,\bar W, \eta, \bar \eta, {\mathbb L}]&=& \int{\rm d}^4 x \rd^2\q\,\cE \Big\{ \eta W -\hf \eta^2 - \frac{1}{4} W^2\Big\} 
+{\rm c.c.} \non \\
&& \quad+ \frac{1}{4} \int \rd^4 x \rd^2 \q \rd^2\bar \q \,E \,
\frac{\eta^2 \bar \eta^2}{{\mathbb L}^2} \fF\Big( \frac{v}{{\mathbb L}^2}, \frac{\bar v}{{\mathbb L}^2}\Big)~.
\eea
This action is obviously super-Weyl invariant.

\section{Duality rotations in $\cN=2$ supersymmetric nonlinear electrodynamics}

Finally,  we give a new realization for the duality-invariant $\cN=2$ supersymmetric 
theories presented in \cite{KT1,K_N=2}. The superspace formulation for $\cN=2$ 
conformal supergravity developed in \cite{KLRT-M} is used throughout  this section.

We denote by $ S[W , {\bar W}]$ an action functional which generates  the dynamics
of an $\cN=2$ vector multiplet. 
The Abelian vector multiplet  coupled to $\cN=2$ conformal supergravity can be described 
by its covariantly chiral field strength $W$, 
\bea
\cDB_{\ad i} W= 0~, 
\label{3}
\eea
  subject to the Bianchi identity
\bea
\Big( \cD^{ij} + 4S^{ij}\Big) W&=&
\Big(\bar \cD^{ij} +  4\bar{S}^{ij}\Big)\bar{W} ~,
\label{n=2bi-i}
\eea
where  $\cD^{ij}:= \cD^{\a(i}\cD_\a^{j)}$ and 
$\bar \cD^{ij} :=  \cDB_\ad{}^{(i}\cDB^{j) \ad}$; 
$S^{ij} $ and its conjugate ${\bar S}_{ij} =\ve_{ik} \ve_{jl} \bar S^{kl}$  
are special dimension-1 components of the torsion, see Appendix B.
In the flat superspace limit, the Bianchi identity reduces to that given in \cite{GSW}.

To realize $W$ as a gauge-invariant field strength, we make use of
a curved-superspace extension 
of Mezincescu's prepotential \cite{Mezincescu} (see also \cite{HST}),  $V_{ij}=V_{ji}$,
which is an unconstrained real SU(2) triplet, $(V_{ij})^* = V^{ij}= \ve^{ik}\ve^{jl}V_{kl}$. 
The expression for $W$ in terms of $V_{ij}$ 
was found in \cite{ButterK} to be
\begin{align}
W = 
\bar\Delta \Big({\cD}^{ij} + 4 S^{ij}\Big) V_{ij}~.
\label{Mez}
\end{align}
Here $\bar\Delta$ is the covariantly chiral projection operator \eqref{chiral-pr}. 

Starting from the action $ S[W , {\bar W}]$, 
we  introduce a covariantly chiral scalar superfield
 $M$ defined as 
\be
{\rm i}\, M := 4\, \frac{\d }{\d W}\,
S[W , {\bar W}]
~,  \qquad \cDB_{\ad i} M= 0~. 
\label{n=2vd}
\ee
In terms of $M$ and its conjugate $\bar M$,  the equation of motion for $V_{ij}$ is
\bea
\Big(\cD^{\a(i}\cD_\a^{j)}+4S^{ij}\Big) M&=&
\Big(\cDB_\ad{}^{(i}\cDB^{j) \ad}+4\bar{S}^{ij}\Big)\bar{M} ~.
\label{n=2em}
\eea
Here we have used the chiral integration rule \eqref{chiral_action_rule}.

Since the Bianchi identity (\ref{n=2bi-i}) and the equation of
motion (\ref{n=2em}) have the same functional form,
one can consider infinitesimal U(1) duality rotations
\be
\d W = \l \, M~, \qquad 
\d M  = -\l \, W~,
\label{n=2dt}
\ee
with $\l$ a constant parameter. The theory under consideration is 
duality invariant under the condition  \cite{K_N=2} 
\bea
{\rm Im} \int \rd^4 x \,{\rm d}^4\q \,\cE\,  \Big( W^2 + M^2 \Big) =0~.
\label{SDE}
 \eea
In the rigid superspace limit, this reduces to the $\cN=2$ self-duality equation \cite{KT1}.

All $\cN=2$ locally supersymmetric theories, which solve the equation \eqref{SDE}, 
are self-dual under a Legendre transformation \cite{KT1,K_N=2}.

At this point, an important comment is in order.
We realize $\cN=2$ Poincar\'e supergravity as a super-Weyl invariant 
coupling of conformal supergravity to certain compensators. 
In such an approach, any matter action, in particular  $ S[W , {\bar W}]$, 
must be  super-Weyl invariant, $ \d_\s  S[W , {\bar W}] =0$.
In the case of duality-invariant theories, 
the self-duality equation \eqref{SDE} has to be super-Weyl invariant. 
Let us check that this is indeed true.
Under the super-Weyl transformation \eqref{super-Weyl}, $W$ varies as \cite{KLRT-M}
\be
\d_{\s} W = \s W~.
\label{Wsuper-Weyl}
\ee
This is induced by 
the following variation of Mezincescu's prepotential  \cite{K_N=2}:
\bea
\d_\s V_{ij} = -(\s +\bar \s) V_{ij}~.
\eea 
Making use of \eqref{Wsuper-Weyl} and 
the super-Weyl transformation of the chiral density
\cite{KLRT-M},
\bea
\d_\s \cE = -2 \s \cE~,
\eea 
we obtain the super-Weyl transformation of $M$:
\bea
\d_{\s} M = \s M~.
\label{Msuper-Weyl}
\eea
Since the chiral scalars $W$ and $M$ have the same super-Weyl transformation law, 
the duality rotation \eqref{n=2dt} is well defined.

We are interested in developing an alternative formulation for the theory described by 
 $ S[W , {\bar W}]$.  For this we consider an auxiliary action of the form
\bea
S[W,\bar W, \eta, \bar \eta]= \hf \int \rd^4 x \rd^4 \q  \,\cE \Big\{ \eta W -\hf \eta^2 - \frac{1}{4} W^2\Big\} 
+{\rm c.c.}  + {\mathfrak S}_{\rm int} [\eta, \bar \eta]~,
\label{4.12}
\eea
in which the scalar superfield $\eta$ is only constrained to be covariantly chiral, $\bar \cD_\bd \eta =0$. 
Here $\fS_{\rm int} [\eta, \bar \eta]$ contains terms of third and higher orders in powers of $\eta$ and its conjugate. We require $S[W,\bar W, \eta, \bar \eta]$ to be super-Weyl invariant, and therefore 
the super-Weyl transformation of $\eta $ is 
\bea
\d_{\s} \eta = \s \eta~.
\label{eta_super-Weyl}
\eea
For $\fS_{\rm int} [\eta, \bar \eta]~$ to be super-Weyl invariant, 
\bea
\d_\s \fS_{\rm int} [\eta, \bar \eta]=0~,
\eea
it may explicitly depend on the supergravity compensators. 
Consider the equation of motion for $\eta$:
\bea
W= \eta- 2\frac{\pa}{\pa \eta}  \fS_{\rm int} [\eta, \bar \eta]~.
\label{4.15}
\eea
In perturbation theory, this equation may be used to express $\eta$ as a functional 
of the field strength $W$ and its conjugate, $\eta =\eta [W, \bar W]$.
As a result, we end up with the action 
\bea
S[W,\bar W] = S[W,\bar W, \eta, \bar \eta]\Big|_{\eta =\eta [W, \bar W]}
=\frac{1}{8} \int \rd^4 x \rd^4 \q  \,\cE \, W^2 
+{\rm c.c.}  + S_{\rm int} [W, \bar W]~,
\label{4.16}
\eea
which describes the dynamics of the vector multiplet. 

Now, we have two expressions for $M$ derived from the actions  \eqref{4.12}
and \eqref{4.16}:
\begin{subequations}
\bea
\ri M &=& -W +2\eta \label{4.17a}\\
&=&  \phantom{-}W + 4 \frac{\d}{\d W} S_{\rm int} [W, \bar W]~.
\label{4.17b}
\eea
\end{subequations}
An important corollary of  eqs.  \eqref{4.17a} and  \eqref{4.17b} is
\bea
\eta = W +  2 \frac{\d}{\d W} S_{\rm int} [W, \bar W]~.
\eea
This relation may be used to determine $W$ as a functional of $\eta$ and 
its conjugate, $W=W[\eta, \bar \eta]$, and then reconstruct the self-interaction 
$\fS_{\rm int} [\eta, \bar \eta]$ starting from the action \eqref{4.16}.

Eq.  \eqref{4.17a} tells us that $\eta =\hf ( W +\ri M)$, 
and hence the duality rotation \eqref{n=2dt} acts on $\eta$ by the rule 
\bea
\d \eta = -\ri \l \eta~.
\label{4.19}
\eea
Making use of the relation \eqref{4.17a} and the equation of motion for $\eta$,  \eqref{4.15}, 
one may see that the self-duality equation \eqref{SDE} is equivalent to 
\bea
{\rm Im} \int \rd^4 x \rd^4 \q  \,\cE \,\eta \frac{\d}{\d \eta} \fS_{\rm int} [\eta, \bar \eta] =0~.
\eea
This means that the self-interaction $\fS_{\rm int} [\eta, \bar \eta] $ is invariant under 
the rigid U(1) transformations \eqref{4.19}. 

Given an arbitrary real functional $\fS_{\rm int} [\eta, \bar \eta]  =\cO(|\eta|^3)$ such that 
\bea
\fS_{\rm int} [ \re^{\ri \vf}  \eta, \re^{-\ri \vf} \bar \eta] = \fS_{\rm int} [\eta, \bar \eta] ~, 
\qquad \vf \in {\mathbb R}~,
\eea
eq.  \eqref{4.12} defines a U(1) duality-invariant  theory. This is the fundamental significance of 
the representation  \eqref{4.12}. 

\section{Concluding comments}
${}$From the point of view of perturbative quantum theory, 
the important features of the new representation 
$S[W,\bar W] \to S[W,\bar W, \eta, \bar \eta]$
are that (i) it can be carried out under a path integral; (ii) it makes the action at most quadratic
in the physical vector multiplet and shifts all self-interaction to the sector of auxiliary chiral variables
$\eta$ and $\bar \eta$. Within the background-field method applied to $S[W,\bar W, \eta, \bar \eta]$, 
we can integrate out the physical vector multiplet without losing duality invariance
(ignoring the issue that models for nonlinear electrodynamics are non-renormalizable).
For a recent  discussion of duality symmetry in perturbative quantum theory see \cite{RoiT}.

The novel representation for duality-invariant $\cN=2$ supersymmetric theories 
presented in section 4 may be useful
for the construction of an $\cN=2$ supersymmetric Born-Infeld action.  
The perturbative scheme to derive such an action was formulated in \cite{KT2}. 
The $\cN=2$ Born-Infeld action should be (i) self-dual; (ii) reduce to the $\cN=1 $ supersymmetric 
Born-Infeld action \cite{CF} under $\cN=2 \to \cN=1$ reduction; and (iii) possess a rigid shift symmetry 
of the form 
\bea
W \to c + \cO( |W|)~, \qquad c \in {\mathbb C}~,
\label{5.1}
\eea
where the field-dependent part should be consistent with (the rigid version of) the Bianchi 
identity \eqref{n=2bi-i}. 
If we work within the representation  \eqref{4.12}, the condition (i) is easy to implement. 
However,  the condition (iii) remains highly non-trivial.  In this formulation, 
the shift  symmetry \eqref{5.1} turns into 
\bea
W \to c + \cO( |\eta|)~,\qquad \eta \to  c + \cO( |\eta|)~. 
\label{5.2}
\eea
It would be really interesting to revisit the problem of constructing the $\cN=2$ Born-Infeld action
by using the representation  \eqref{4.12}.

In conclusion, we wish to emphasize that the approaches developed \cite{KT2,KT1} and 
\cite{KMcC,K_N=2} 
apply to arbitrary rigid and locally supersymmetric duality invariant theories. This is in stark 
contradistinction with the bosonic approaches of \cite{GR1,GZ2,GZ3} which literally apply only to the theories without higher derivatives, i.e. when the Lagrangian does not involve derivatives of the field strengths, 
$L= L(F_{ab})$. An extension of the formalism  of \cite{GR1,GZ2,GZ3}
to the case of $\cN=0$ duality invariant theories 
with higher derivatives trivially follows from the $\cN=1$ supersymmetric approach of \cite{KT1}.
To derive such an extension, one may start from  the rigid version 
of the $\cN=1$ supersymmetric self-duality equation \eqref{2.6}
and switch off all the fermionic and auxiliary fields. This will lead to the following  
self-duality equation for an electromagnetic field theory with action $S[F]$
\bea
\int \rd^4x\, \Big\{G^{ab} \tilde{G}_{ab} +F^{ab}  \tilde{F}_{ab} \Big\}= 0~,
\label{5.3}
\eea
where we have defined \cite{KT2}
\bea
\tilde G_{ab} [F] ~=~2 \,\frac{\delta S[F] }{ \delta F^{ab}}~.
\label{5.4}
\eea
In these relations $F_{ab}$ has to chosen to be an arbitrary bivector, $F_{ab}=-F_{ba}$, 
which is not subject to the Bianchi identity.  In the case of theories without higher derivatives, 
$S[F] = \int \rd^4x\, L(F)$, the self-duality equation \eqref{5.3} is equivalent to \eqref{GR}.
Ref. \cite{KT2} clearly sketched the steps leading to the equation \eqref{5.3}
(at the end of section 2),
although this equation was 
not given explicitly. It appears that  it had escaped the attention of the authors of 
 the recent publication \cite{Chemissany:2011yv} 
that the procedure which leads to eq. \eqref{5.3} was already outlined in \cite{KT2}.
\\

\noindent
{\bf Acknowledgements:}\\
Email correspondence with Evgeny Ivanov and Boris Zupnik is gratefully acknowledged.  The author is  grateful to Stefan Theisen for comments on the manuscript. It is a pleasure to thank Joseph Novak for reading the manuscript. 
This work was supported in part by the Australian Research Council.

\appendix

\section{The Wess-Zumino superspace geometry}

The superspace geometry is described 
by covariant derivatives of the form
\bea
\cD_A &=& (\cD_a , \cD_\a ,\cDB^\ad ) = E_A + \O_A~.
\eea
Here $E_A$ denotes the inverse vielbein, 
$E_A = E_A{}^M  \pa_M $,
and $\O_A$  the Lorentz connection, 
$
\O_A = \O_A{}^{\b \g} M_{\b \g}
+\O_A{}^{\bd \gd} {\bar M}_{\bd \gd} $,
with  $M_{\b\g}$ and $ {\bar M}_{\bd \gd}$
the Lorentz generators.
The covariant derivatives obey the following anti-commutation relations:
\begin{subequations}\label{algebra}
\bea
  \{ \cD_\a , {\bar \cD}_\ad \} &=& -2{\rm i} \cD_{\a \ad} ~,  \\
\{\cD_\a, \cD_\b \} &=& -4{\bar R} M_{\a \b}~, \qquad
\{ {\bar \cD}_\ad, {\bar \cD}_\bd \} =  4R {\bar M}_{\ad \bd}~,  \\
\left[ { \bar \cD}_{\ad} , \cD_{ \b \bd } \right]
     & = & -{\rm i}{\ve}_{\ad \bd}
\Big(R\,\cD_\b + G_\b{}^{\dot{\g}}  \cDB_{\dot{\g}}
-(\cDB^\gd G_\b{}^{\dot{\d}})
{\bar M}_{\gd \dot{\d}}
+2W_\b{}^{\g \d}
M_{\g \d} \Big)
- {\rm i} (\cD_\b R)  {\bar M}_{\ad \bd}~,~~~~~~~  \\
\left[ \cD_{\a} , \cD_{ \b \bd } \right]
     & = &
     {\rm i}
     {\ve}_{\a \b}
\Big({\bar R}\,\cDB_\bd + G^\g{}_\bd \cD_\g
- (\cD^\g G^\d{}_\bd)  M_{\g \d}
+2{\bar W}_\bd{}^{\gd \dot{\d}}
{\bar M}_{\gd \dot{\d} }  \Big)
+ {\rm i} (\cDB_\bd {\bar R})  M_{\a \b}~. 
\eea
\end{subequations}
Here the torsion tensors $R$, $G_a = {\bar G}_a$ and
$W_{\a \b \g} = W_{(\a \b\g)}$ satisfy certain Bianchi identities \cite{BK,WB}.
In particular, $R$ and $W_{\a\b\g}$ are covariantly chiral. 

The chiral integration rule in $\cN=1$ supergravity is 
\bea
\int \rd^4 x \,{\rm d}^2\q {\rm d}^2{\bar \q}\,E\, U 
=  -\frac{1}{4} \int \rd^4 x \,{\rm d}^2\q \,
\,\cE\, (\bar \cD^2 -4R) U 
~, \qquad E^{-1}= {\rm Ber}(E_A{}^M)~,
\label{chiral_action_rule_N=1}
\eea
with $\cE$ the chiral density. 

\section{$\cN = 2$ conformal supergravity} 

In this appendix we give a summary of the  superspace formulation for $\cN=2$ conformal supergravity
developed in \cite{KLRT-M}.

The structure group is chosen to be $\rm SL(2, \dsC) \times SU(2)$, 
and the covariant derivatives $\cD_A = (\cD_a, \cD_\a^i, \bar \cD^\ad_i)$
read
\bea
\cD_A &=& E_A + \Phi_A{}^{kl} J_{kl}+ \Omega_A{}^{\b\g} M_{\b\g} 
  + \bar{\Omega}_A{}^{ \dot{\b} \dot{\g} } \bar M_{\dot{\b}\dot{\g}}~.
\eea
Here $J_{kl}$ are the generators of the group SU(2), 
and $\Phi_A{}^{kl}$ the corresponding connection.

The spinor covariant derivatives obey the anti-commutation relations \cite{KLRT-M}
\begin{subequations} 
\bea
\{\cD_\a^i,\cD_\b^j\}&=&\phantom{+}
4S^{ij}M_{\a\b}
+2\ve^{ij}\ve_{\a\b}Y^{\g\d}M_{\g\d}
+2\ve^{ij}\ve_{\a\b}\bar{W}^{\gd\dd}\bar{M}_{\gd\dd}
\non\\
&&
+2 \ve_{\a\b}\ve^{ij}S^{kl}J_{kl}
+4 Y_{\a\b}J^{ij}~,
\label{acr1} \\
\{\cDB^\ad_i,\cDB^\bd_j\}&=&
-4\bar{S}_{ij}\bar{M}^{\ad\bd}
-2\ve_{ij}\ve^{\ad\bd}\bar{Y}^{\gd\dd}\bar{M}_{\gd\dd}
-2\ve_{ij}\ve^{\ad\bd}{W}^{\g\d}M_{\g\d} \non \\
&&
-2\ve_{ij}\ve^{\ad\bd}\bar{S}^{kl}J_{kl}
-4\bar{Y}^{\ad\bd}J_{ij}~,
\label{acr2} \\
\{\cD_\a^i,\cDB^\bd_j\}&=&
-2\ri\d^i_j(\s^c)_\a{}^\bd\cD_c
+4\d^{i}_{j}G^{\d\bd}M_{\a\d}
+4\d^{i}_{j}G_{\a\gd}\bar{M}^{\gd\bd}
+8 G_\a{}^\bd J^{i}{}_{j}~.
\label{acr3}
\eea
\end{subequations}
Here the real four-vector $G_{\a \ad} $,
the complex symmetric  tensors $S^{ij}=S^{ji}$, $W_{\a\b}=W_{\b\a}$, 
$Y_{\a\b}=Y_{\b\a}$ and their complex conjugates 
$\bar{S}_{ij}:=\overline{S^{ij}}$, $\bar{W}_{\ad\bd}:=\overline{W_{\a\b}}$,
$\bar{Y}_{\ad\bd}:=\overline{Y_{\a\b}}$ obey additional differential constraints implied 
by the Bianchi identities \cite{Grimm,KLRT-M}.

An infinitesimal  super-Weyl transformation of the covariant derivatives \cite{KLRT-M} is
\begin{subequations} \label{super-Weyl} 
\bea
\d_{\s} \cD_\a^i&=&\hf\sba\cD_\a^i+(\cD^{\g i}\s)M_{\g\a}-(\cD_{\a k}\s)J^{ki}~, \label{super-Weyl1} \\
\d_{\s} \cDB_{\ad i}&=&\hf\s\cDB_{\ad i}+(\cDB^{\gd}_{i}\sba)\bar{M}_{\gd\ad}
+(\cDB_{\ad}^{k}\sba)J_{ki}~, 
\label{super-Weyl2} 
\eea
\end{subequations}
where  the parameter $\s$ is an arbitrary covariantly chiral superfield, $\bar \cD^\ad_i \s =0$.

The covariantly chiral projection operator \cite{Muller} is 
\bea
\bar{\D}
&=&\frac{1}{96} \Big((\cDB^{ij}+16\bar{S}^{ij})\cDB_{ij}
-(\cDB^{\ad\bd}-16\bar{Y}^{\ad\bd})\cDB_{\ad\bd} \Big)
\non\\
&=&\frac{1}{96} \Big(\cDB_{ij}(\cDB^{ij}+16\bar{S}^{ij})
-\cDB_{\ad\bd}(\cDB^{\ad\bd}-16\bar{Y}^{\ad\bd}) \Big)~,
\label{chiral-pr}
\eea
where $\cDB^{\ad\bd}:=\cDB^{(\ad}_k\cDB^{\bd)k}$. 
The  fundamental property  of $\bar \D$ is that $\bar{\D} U$ is covariantly chiral,
for any scalar and isoscalar superfield $U$,
that is ${\bar \cD}^{\ad}_i \bar{\D} U =0$. 
This operator relates an integral over the full superspace to that over its chiral subspace:
\bea
\int \rd^4 x \,{\rm d}^4\q \,{\rm d}^4{\bar \q}\,E\, U 
=  \int \rd^4 x \,{\rm d}^4\q \,
\,\cE\, \bar\D U 
~, \qquad E^{-1}= {\rm Ber}(E_A{}^M)~,
\label{chiral_action_rule}
\eea
with $\cE$ the chiral density.
A   derivation of \eqref{chiral_action_rule} is given \cite{KT-M}.

\begin{footnotesize}

\end{footnotesize}

\end{document}